%% file: DLSR_TMM_main.tex
\documentclass[lettersize,journal]{IEEEtran}
\IEEEoverridecommandlockouts
\usepackage{amsmath,amsfonts,amssymb}
\usepackage{array}
\usepackage[caption=false,font=normalsize,labelfont=sf,textfont=sf]{subfig}
\usepackage{textcomp}
\usepackage{stfloats}
\usepackage{url}
\usepackage{verbatim}
\usepackage{graphicx}
\usepackage{cite}
\usepackage{times}
\usepackage{multirow}
\usepackage{epsfig}
\usepackage{color}
\usepackage{lipsum}
\newcommand{\etal}{\textit{et al}.}
\usepackage[linesnumbered,ruled,vlined]{algorithm2e}
\begin{document}

\title{Differentiable Neural Architecture Search for Extremely Lightweight Image Super-Resolution}

\author{Han Huang, Li Shen, Chaoyang He, Weisheng Dong, and Wei Liu~\IEEEmembership{Fellow,~IEEE}

\thanks{This work is supported by the Major Science and Technology Innovation 2030 “Brain Science and Brain-like Research” key project (No. 2021ZD0201405)). \emph{(Corresponding author: Li Shen)}}

\thanks{Han Huang and Wei Liu are with Tencent, Shenzhen 518057, China.}
\thanks{Li Shen is with JD Explore Academy, Beijing, China (e-mail: mathshenli@gmail.com)}
\thanks{Chaoyang He is with the Department of Computer Science, University of Southern California, Los Angeles, CA 90007 USA.}
\thanks{Weisheng Dong is with School of Artificial Intelligence, Xidian University, Xi’an, China.}
\thanks{Manuscript received November 19, 2022; accepted December 11, 2022}}

\markboth{Journal of \LaTeX\ Class Files,~Vol.~14, No.~8, August~2021}%
{Shell \MakeLowercase{\textit{et al.}}: A Sample Article Using IEEEtran.cls for IEEE Journals}


\maketitle

\IEEEpubid{\begin{minipage}{\textwidth}\ \\[12pt] \centering
  Copyright © 2022 IEEE. Personal use of this material is permitted. However, permission to use this material \\
  for any other purposes must be obtained from the IEEE by sending an email to pubs-permissions@ieee.org.
\end{minipage}} 



\IEEEpubidadjcol


\input{1.abstract}
\input{2.Introduction}
\input{3.Related_Work}
\input{4.Method}

\input{5.Experiment}

\input{6.Conclusion}

 \bibliographystyle{IEEEtran}
 \bibliography{IEEEabrv, reference}

\end{document}

%% file: 1.abstract.tex
\begin{abstract}
  Single Image Super-Resolution (SISR) tasks have achieved significant performance with deep neural networks. 
  However, the large number of parameters in CNN-based met-hods for SISR tasks require heavy computations. 
  Although several efficient SISR models have been recently proposed, most are handcrafted and thus lack flexibility. In this work, we propose a novel differentiable Neural Architecture Search (NAS) approach on both the cell-level and network-level to search for lightweight SISR models. Specifically, the cell-level search space is designed based on an information distillation mechanism, focusing on the combinations of lightweight operations and aiming to build a more lightweight and accurate SR structure. The network-level search space is designed to consider the feature connections among the cells and aims to find which information flow benefits the cell most to boost the performance. 
  Unlike the existing Reinforcement Learning (RL) or Evolutionary Algorithm (EA) based NAS methods for SISR tasks, our search pipeline is fully differentiable, and the lightweight SISR models can be efficiently searched on both the cell-level and network-level jointly on a single GPU.
  Experiments show that our methods can achieve state-of-the-art performance on the benchmark datasets in terms of PSNR, SSIM, and model complexity with merely 68G Multi-Adds for $\times 2$ and 18G Multi-Adds for  $\times 4$ SR tasks.
 
\end{abstract}


%% file: 2.Introduction.tex
\section{Introduction}

\IEEEPARstart{I}{mage} super-resolution (SR) is a low-level vision problem that reconstructs a single low-resolution (LR) image to a high-resolution (HR) image. This problem is ill-posed since multiple HR images can degrade to the same LR image.
Many deep-learning-based methods have been proposed to address this problem \cite{dong2015image,Kim_2016_CVPR,Lim_2017_CVPR_Workshops,ledig2017photo,zhang2018image,zhang2018residual,guo2020closed,8708220,ma2020structure, liu2020residual} and have achieved great success.

\begin{figure}[ht]
\centering
\includegraphics[width=0.45\textwidth]{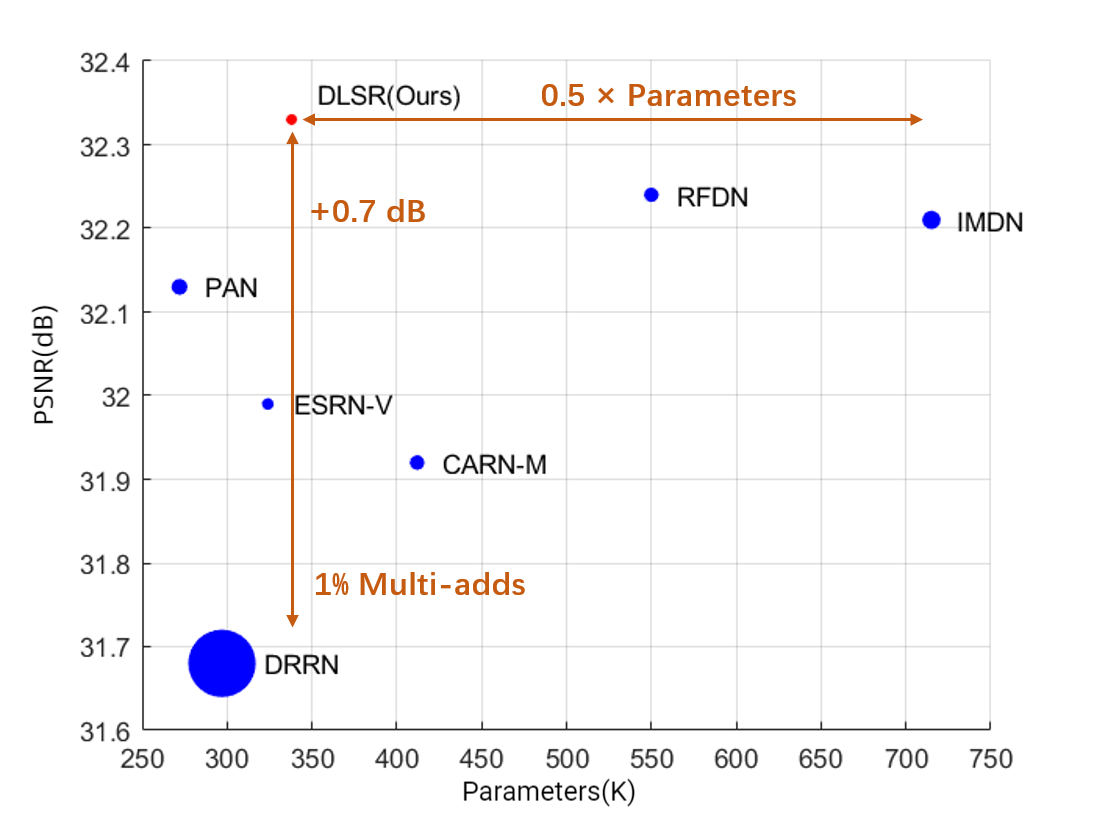}
\caption{Performance comparison of existing lightweight methods on Set5 \cite{bevilacqua2012low} (4$\times$). The size of the dot denotes the Multi-Adds of the method. Our method achieve state-of-the-art performance with fewer parameters or fewer Multi-Adds.}
\label{fig:scatter}
\vspace{-0.3cm}
\end{figure}

While the SR performance is boosted by the deep learning approach, the model complexity is also increased.  For example, RDN \cite{zhang2018residual} had 22M parameters and EDSR \cite{Lim_2017_CVPR_Workshops} reached up to 43M parameters. 
It is difficult to deploy these models to the equipment with low computing power. 
For real-world applications, lightweight and efficient SR models have also been designed in recent years, including handcrafted SR neural networks \cite{kim2016deeply,ahn2018fast,9427111,8672189} and neural architecture search (NAS) based SR methods \cite{9413080,song2020efficient,lee2020journey, pan2020real}.

\IEEEpubidadjcol
Although great improvements have been achieved by existing lightweight SR methods, they still suffer from several limitations.
First, hand-crafted lightweight SR models like IMDN \cite{Hui-IMDN-2019} and RFDN \cite{10.1007/978-3-030-67070-2_2} adopted several $3 \times 3$ convolution layers with large amount of parameters and Multi-Adds.
The building blocks designed by these methods with the same three $3 \times 3$ convolution layers can also be suboptimal and lack flexibility for SISR tasks. 
Second, the network-level architecture of these methods only considered concatenating the output features of the blocks at the end of the model while omitting intermediate information flows among the blocks, which have been demonstrated to enlarge the reception field \cite{huang2017densely} and could be useful for improving SR performance \cite{tong2017image, seif2018large, zhang2018residual, shang2020perceptual}. 
However, the network cannot be connected too densely in order to achieve an efficient lightweight SR model.
Therefore, it's important to find which connection benefits the cell most to improve the performance of lightweight SR models while keeping a slightly low model complexity.
Finally, as a dense prediction task, SR requires to predict the value of each pixel for HR image, which is sensitive to the tiny changes of the architectures of the network. Hence it is challenging to design or search for a suitable network for the SR task. Moreover, the dimensions of the data and the model of the SR task are much larger than those of the classification task, which is more challenging to search in a differentiable manner. Most Neural Architecture Search (NAS) based methods for SR tasks were based on reinforcement learning and evolutionary methods which are time-consuming and require a large number of computing resources to search for appropriate models. Furthermore, they failed to achieve better peak signal-to-noise ratio (PSNR) or structural similarity index measure (SSIM) \cite{wang2004image} results with searched lightweight SR models comparing with the existing state-of-the-art (SOTA) hand-crafted SR models.

To address these problems, we propose a lightweight image super-resolution method with a fully differentiable neural architecture search (DLSR) which is composed of cell-level and network-level search techniques. 
For cell-level search, we design a large search space (see Table \ref{tab:Operations}) that contains more lightweight convolution operations to increase the probabilities of finding more lightweight models.
As opposed to existing work \cite{liu2018darts, song2020efficient} that searched for arbitrary combinations and connections of basic operations or searched for handcrafted blocks, we search for the operation combinations based on information distillation structure that provides prior knowledge of nice lightweight SR structures.  
Based on the flexibility of lightweight operation combinations, our search space not only contains the handcrafted RFDB \cite{10.1007/978-3-030-67070-2_2} structure but also explores for a better cell for efficient SR.  
To utilize the intermediate information flow between the cells, we design a network-level search space that contains all possible connections among the cells to further boost the performance. 
As opposed to FALSR \cite{9413080} which uses an evolutionary algorithm to search for block connections with discrete encoding, we first densely connect the blocks to build a super-net, then utilize the continuous relaxed architecture parameters to weigh the connections and optimize the parameters with the stochastic gradient descent method. 
During searching, the network automatically identifies the most important intermediate information flow connections. 

In addition, we design a loss function composed of three parts: L1 loss, Hign Frequency Error Norm (HFEN) loss \cite{ravishankar2010mr}, and the number of parameters of the operations. HFEN is an image comparison metric from medical imaging and uses a Laplacian of Gaussian kernel for edge-detection. Thus, the HFEN loss can help to minimize the reconstruction error of high-frequency image details. 
In addition, we treat the number of parameters of the operations as a regularization term to push the searching direction into a more lightweight space. Experimental results show that our DLSR method surpasses other SOTA lightweight SR methods in terms of PSNR, SSIM with fewer parameters and Multi-Adds on benchmark datasets: Set5 \cite{bevilacqua2012low}, Set14 \cite{yang2010image}, B100 \cite{martin2001database}, and
Urban100 \cite{huang2015single} in $\times2, \times 3, \times 4 $ super-resolution tasks.
Moreover, the differentiable NAS strategy we adopted enables the optimization procedure for bi-level search simultaneously on single GPU. 
In the end, our main contributions are summarized as three-fold:
\begin{itemize}
\item We propose a differentiable NAS strategy for searching a lightweight SR model, which incorporates both cell-level and network-level search spaces to strengthen the SR performance. The proposed approach significantly reduces the searching cost compared to existing RL-based NAS methods. 
\item We design a loss function that jointly considers distortion, high-frequency reconstruction, and lightweight regularization that push the searching direction to explore a better lightweight SR model. 
\item We conduct extensive experiments to evaluate the efficacy of our method, which achieves state-of-the-art performance on the benchmark datasets in terms of PSNR, SSIM, and model complexity.
\end{itemize}

%% file: 3.Related_Work.tex
\section{Related work}

\subsection{CNN-based Image Super-Resolution}
 SR performance has been greatly improved by CNN-based methods \cite{dong2015image, Kim_2016_CVPR, tai2017memnet, Lim_2017_CVPR_Workshops, ledig2017photo, zhang2018image, zhang2018residual, wang2018esrgan, zhang2018learning,8708220,zhang2020deep, liu2020residual}. Dong \etal\cite{dong2015image} propose SRCNN which is a shallow three-layer network to map interpolated LR images to HR images.
 Kim \etal\cite{Kim_2016_CVPR} propose the VDSR network, which is composed of 20 layers and global skip-connection to improve the performance.
In addition, Dong \etal\cite{dong2016accelerating} design the transposed convolution layer, and Shi \etal\cite{shi2016real} propose the sub-pixel convolution layer for SR tasks. Both methods perform the upsampling operation at the end of the CNN, hence largely save on computation in the feature extraction phase due to the reduction of spatial dimension. Lim \etal \cite{Lim_2017_CVPR_Workshops} propose EDSR and MDSR, which remove Batch Normalization layers in SRResnet  \cite{ledig2017photo} and greatly improve the performance.
Zhang \etal\cite{zhang2018residual} propose RDN networks by introducing dense connections into EDSR residual blocks. RCAN \cite{zhang2018image} introduces channel attention to achieve better SR performance. However, most of these CNN-based methods contain large parameters and require large amounts of computation, which limits their real-world applications.

\begin{figure*}[t]
\centering
\subfloat[The architecture of cell]{
\includegraphics[width=0.45\textwidth]{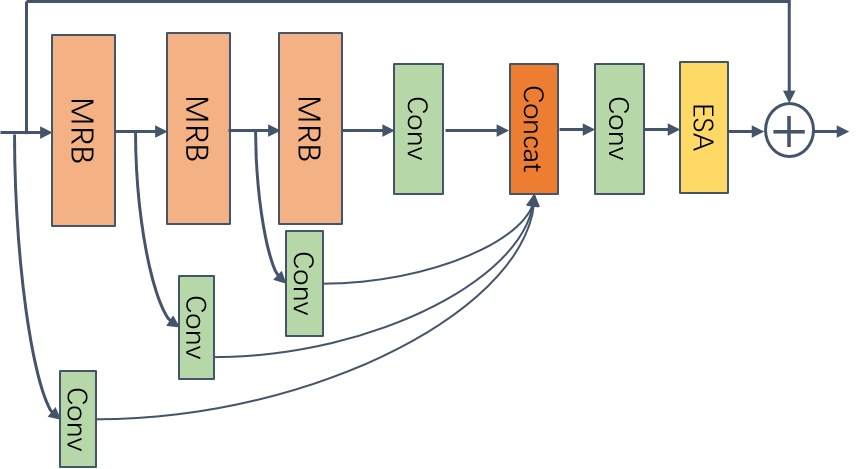}
}
\hspace{1cm}
\subfloat[The architecture of MRB]{
\includegraphics[width=0.25\textwidth]{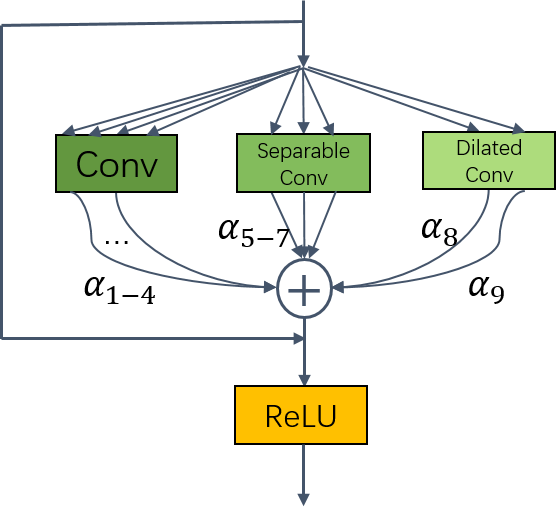}
}

\caption{The cell-level search space. The cell is composed of 3 mixed residual blocks with an information distillation mechanism and an ESA block. The `Conv' in figure(a) denotes the $1\times 1$ convolution layer that cuts the channel number by half. Figure(b) shows the architecture of mixed block, which is composed of multiple operations weighted by parameter $\alpha$, residual skip connection, and ReLU layer.}
\label{fig:cell}
\vspace{-0.3cm}
\end{figure*}

\begin{figure*}
\begin{center}
\includegraphics[width=0.55\linewidth]{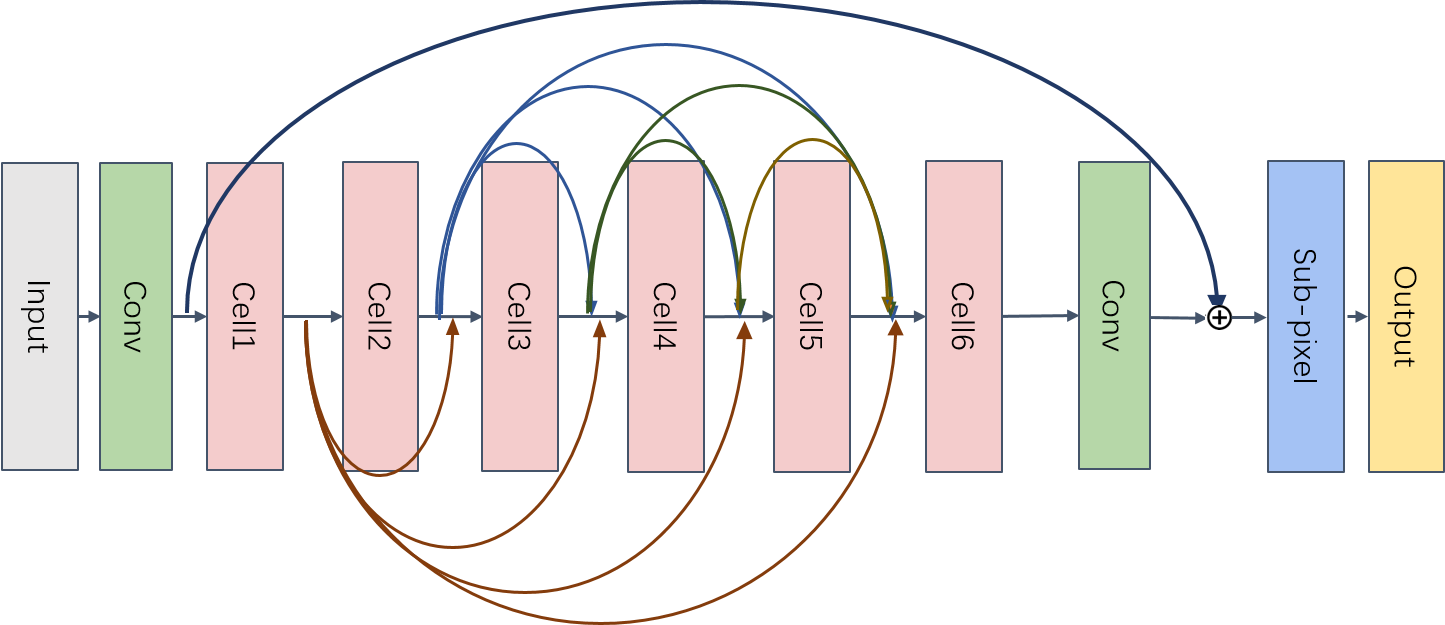}
\end{center}
\vspace{-0.3cm}
   \caption{The network-level search space. Each cell is connected with all of its prior cells. And each connection is weighted by architecture parameter $\beta$. Each cell's input feature is composed of weighted features from the prior cells through concatenation and $1 \times 1$ convolution. The connections from each cell to the last convolution layer are omitted for clarity.}
\label{fig:network-level}
\vspace{-0.3cm}
\end{figure*}

\subsection{Lightweight Image Super-Resolution}
Lightweight and efficient CNN for SR tasks has been widely explored to suit mobile devices with an extremely small amo-unt of parameters and computation \cite{tai2017image, ahn2018fast, hui2018fast, Hui-IMDN-2019, zhu2019efficient, 10.1007/978-3-030-58542-6_17, zhang2020aim, 10.1007/978-3-030-67070-2_2,9427111}.
In order to reduce the parameters, Kim \etal\cite{tai2017image} introduce recursive layers combined with residual schemes in the feature extraction stage.
Ahn \etal\cite{ahn2018fast} propose the CARN-M model which utilizes group convolution and cascade network architecture that significantly reduces the parameters. 
Hui \etal\cite{hui2018fast,Hui-IMDN-2019}  propose an information distillation mechanism (IDM)  that utilizes a channel splitting strategy to distill and compress local short-path feature information.
RFDN \cite{10.1007/978-3-030-67070-2_2} rethinks the channel splitting strategy and decouples the convolution layer and channel splitting layer.
Furthermore, they apply the skip-connection on the $3 \times 3$ convolution that makes up the shallow residual block (SRB), which significantly improves the SR performance. 
Pan \etal\cite{pan2022dual} propose DualCNN which has two parallel branches and respectively recovers the structures and details in an end-to-end manner.

\subsection{SISR with Neural Architecture Search}
As NAS techniques have achieved great success in image classification \cite{liu2018darts,he2020milenas} and other tasks, recent works have started to adopt NAS to search for efficient SR networks.
FALSR \cite{9413080} utilizes reinforcement learning and evolutionary methods to search for lightweight SR models, building SR as a constrained multi-objective optimization problem.
Song \etal  \cite{song2020efficient} propose to search for multiple handcrafted efficient residual dense blocks to stack the SR model using evolutionary methods.
Guo \etal \cite{guo2020hierarchical} propose to search for cell structures and upsampling positions with reinforcement learning.
Recently, TPSR \cite{lee2020journey} has adopted reinforcement learning to find an efficient GAN-based SR model, resulting a tiny  SR model that performes well on both perceptual and distortion metrics.
However, most of the prior SR methods with NAS utilized reinforcement learning or evolutionary methods that were time-consuming. In this work, we explore the fully differentiable NAS to search for an efficient, accurate, and lightweight SR model with a single GPU.

\subsection{Related work in IEEE TCSVT}

There are several papers published in the IEEE Transactions on Circuits and Systems for Video Technology that are
most closely related to our work. CSFM \cite{8708220} introduces Channel-wise and spatial attention to capture more informative
features. However, the large number of parameters of the model require heavy computations. FilterNet\cite{8672189} presents the dilated residual group which adaptively filters the redundant low-frequency information. EMASRN\cite{9427111} proposes expectation-maximization attention mechanism for better balancing performance and applicability. However, these methods are handcrafted and thus lack flexibility. We propose a differentiable NAS strategy for searching a lightweight SR model, which incorporates both cell-level and network-level search spaces to strengthen the SR performance. We design a loss function that jointly considers distortion, high-frequency reconstruction, and lightweight regularization that push the searching direction to explore a better lightweight SR model.

%% file: 4.Method.tex
\section{Method}

In this section, we introduce our {\bf D}ifferentiable NAS method for {\bf L}ightweight {\bf S}uper-{\bf R}esolution model, dubbed DLSR. 
Below, we first describe the search space of cell-level and network-level. Then, we discuss the search strategy and loss function of our proposed DLSR.

\subsection{Search Space}
{\bf Cell-level search space} The cell-level topology structure is based on residual feature distillation block (RFDB) \cite{10.1007/978-3-030-67070-2_2}, which is comprised of three shallow residual blocks (SRB) with an information distillation mechanism and a contrast-aware channel attention (CCA) layer \cite{Hui-IMDN-2019}. 
The smallest building block SRB is composed of a $3 \times 3$ convolution layer and a residual connection.
However, we argue that the $3 \times 3$ convolution in RFDB could be suboptimal and would thus not always be the best choice for lightweight super-resolution. In order to improve the flexibility of the RFDB and search for a more lightweight structure, we replace the SRB with {\bf M}ixed {\bf R}esidual {\bf B}lock (MRB) in Figure \ref{fig:cell}. 
The MRB is composed of a mixed layer, a residual connection, and a ReLU layer, in which the mixed layer is made up of multiple operations including separable convolution, dilated convolution, and normal convolution.
For mixed layer $k$, we denote the input feature as {\small $x_k$}, and the operation space as $O$, where each element represents a candidate function $o(\cdot)$ weighted by the cell architecture parameters {\small $\alpha _o^k$}, as illustrated in Figure \ref{fig:cell}. We use softmax to perform the continuous relaxation of the operation space as done in DARTS \cite{liu2018darts}. Thus, the output of mixed layer $k$ denoted by {\small${f_k}({x_k})$} is given as:
\begin{equation}\label{eq:mix_layer}
{f_k}({x_k}) = \sum\limits_{o \in {\rm O}} {\frac{{\exp (\alpha _o^k)}}{{\sum\limits_{o' \in {\rm O}} {\exp (\alpha _{o'}^k)} }}o({x^k})}.
\end{equation}
During searching, the operation with the largest {\small$\alpha _o^k$} is reserved as the genotype of the layer. 
The structure of each cell is composed of three MRBs with a feature distillation mechanism and an enhanced spatial attention (ESA) block as shown in Figure \ref{fig:cell}. Hence, the number of possible combinations for each cell is $9 \times 9 \times 9$. 

\begin{table}[t]
\begin{small}
\begin{center}
\caption{Operations and their complexities in mixed layer. Muti-Adds are calculated in $\times 2$ SR task with 50 channels on 1280×720 image. Dilated convolution \cite{yu2017dilated} is joint with group convolution.}
\begin{tabular}{|c|c|c|c|}
\hline
\multirow{1}{5em}{Operation}& Kernel Size  & Params (K) & Muti-Adds (G) \\
\hline
\hline
\multirow{4}{5em}{convolution}&$1 \times 1$&2.5&0.576 \\
& $3 \times 3$&22.5&5.184\\
& $5 \times 5$&62.5&14.400\\
& $7 \times 7$&122.5&28.224\\
\hline
\multirow{3}{5em}{Separable convolution}\!\!& $3 \times 3$&5.9&1.359\\
&$5 \times 5$&7.5&1.728\\
&$7 \times 7$&9.9&2.281\\
\hline
\multirow{2}{5em}{Dilated convolution}\!\!& $3 \times 3$&2.95&0.680\\
&$5 \times 5$&3.75&0.864\\
\hline
\end{tabular}
\end{center}
\label{tab:Operations}
\end{small}
\vspace{-0.5cm}
\end{table}

\smallskip 
\noindent
{\bf Network-level search space}\ Different from HNAS \cite{guo2020hierarchical} which designs the network-level search space to search for the upsampling positions or HiNAS \cite{zhang2020memory} that is designed to search for the network width, we design the network-level search space to search for the shortcut connections among the cells to explore the intermediate information, as shown in Figure \ref{fig:network-level}. 
The whole network is stacked with $6$ cells, and each cell is connected with all of its predecessors. The output features of its prior cells are concatenated and passed into $1 \times 1$ convolution layer to aggregate the information. In addition, each cell's output feature is connected to the last convolution layer. The connection between cell $i$ and cell $j$, which is also the feature maps of cell $i$, is denoted {\small ${x^{i}}$}, weighted by the network-level architecture parameters {\small ${\beta ^{(i,j)}}$}. We also utilize softmax function as continuous relaxation for parameters {\small ${\beta ^{(i,j)}}$} as Eq.~\eqref{eq:mix_layer}. Then, the input of the cell $j$ denoted by {\small ${I_j}$} is formulated as:

\begin{small}
\begin{equation}
{I_j} = g\left(\frac{{\exp ({\beta ^{(i,j)}})}}{{\sum\limits_{i' < j} {\exp ({\beta ^{(i',j)}})} }}{x^i}\right),
\label{eq:connection}
\end{equation}
\end{small}
\!\!where {\small$g(\cdot)$} denotes the operations of concatenation and $1 \times 1$ convolution. 
Thus, we build a continuous and dense super-network search space considering all the intermediate information among the cells. 

\smallskip 
\noindent
{\bf Search complexity}\ 
Based on the above illustration, the proposed DLSR method includes both the cell-level and network-level search spaces. Thus, the overall search complexity of our method is estimated as:
\begin{equation}\label{eq:search_space}
9 \times 9 \times 9 \times 5 \times 4 \times 3 \times2 = 87480.
\end{equation}

It is nontrivial and requires a large amount of computation cost to explore such a large search space for a lightweight and accurate super-resolution model via reinforcement learning \cite{guo2020hierarchical} or evolutionary algorithm \cite{9413080,song2020efficient} based neural architecture search approaches. 
In this work, we solve this problem via a fully differentiable neural architecture search approach.    

\subsection{Search Strategy}
We extend the popular differentiable NAS methods including DARTS \cite{liu2018darts} and its improved version MiLeNas \cite{he2020milenas} for the low-level computer vision (SISR) task.  
These two methods were originally proposed for the image classification task which is a high-level computer vision task. 
Motivated by MiLeNas, the objective function of our DLSR model is defined as the following regularized form:
\begin{equation}
{\min _{\theta,\alpha ,\beta }}[{L_{tr}}({\theta^*}(\alpha ,\beta ) + \lambda {L_{val}}({\theta^*}(\alpha ,\beta );\alpha ,\beta )],
\label{eq:object_function}
\end{equation}
where {\small$L_{tr}$} denotes the loss on training dataset and {\small$L_{val}$} denotes the loss on validation dataset. The {\small$\theta$} denotes the weights parameters of the network and {\small$\lambda $} is a non-negative regularization parameter that balances the importance of the training loss and validation loss. Because the architecture parameters {\small$\alpha$} and {\small$\beta$} are both continuous, we directly apply Adam \cite{kingma2014adam} to solve problem \eqref{eq:object_function}. We define the architecture parameter {\small$A = [ \alpha, \beta ]$}, the parameters {\small$\theta$}, {\small$\alpha$}, and {\small$\beta$} are updated via the following iteration:
\begin{gather}
    \theta = \theta - {\eta _\theta }{\nabla _\theta}{L_{tr}}(\theta,A); \label{eq:theta_update}\\
    A = A - {\eta _A }({\nabla _A}{L_{tr}}(\theta,A)+\lambda{\nabla _A}{L_{val}}(\theta,A)). \label{eq:A_update}
\end{gather}
During the searching process, we preserve the operation that has the maximal value of the {\small$\alpha$} as the searched operation of the layer. The connections that have the maximal and the submaximal value of the {\small$\beta$} are preserved as the searched input connections of the cell. Our searching and training procedure is summarized in Algorithm \ref{alg:overall}.

\subsection{Loss Function}
To achieve lightweight and accurate SR models, the loss function is composed of three parts, which include L1 loss as distortion loss, HFEN loss \cite{ravishankar2010mr} for reconstruction, and parameters of the operations as a lightweight limitation.
\begin{gather}
L_1 = \frac{1}{N}\sum_{i=1}^{N}\left|({H_\theta }({I^{LR}}) - {I^{HR}})\right| \label{eq:l1_loss}\\
L_{HFEN} = \frac{1}{N}\sum_{i=1}^{N}\left|\nabla{H_\theta(I}^{LR})-\nabla\ I^{HR}\right| \label{eq: HFEN_loss}\\
L_P=\sum_{o\in O}\frac{p_o}{\sum_{c\in O}p_c}{\frac{{\exp (\alpha _o)}}{{\sum_{o' \in {\rm O}} {\exp (\alpha _{o'})} }}}.   \label{eq: latency_loss} \\
L(\theta ) = {L_1} + \mu  \times {L_{HFEN}} + \gamma  \times {L_{P}}  \label{eq:loss_function}. 
\end{gather}
Specifically, {\small$L_1$} loss is popularly used for SR tasks \cite{Lim_2017_CVPR_Workshops,9413080,10.1007/978-3-030-67070-2_2,10.1007/978-3-030-67070-2_3} to minimize the distortion between the reconstructed SR image and ground truth HR image;
{\small$L_{HFEN}$} \cite{chaitanya2017interactive} is a gradient-domain L1 loss, and each gradient {\small$\nabla(\cdot)$} is computed using High Frequency Error Norm (HFEN) \cite{ravishankar2010mr} which is an image comparison metric from medical imaging and uses a Laplacian of Gaussian kernel for edge-detection with Gaussian filter pre-smoothed images. {\small$L_{HFEN}$} is adopted to strengthen the reconstruction of image details such as edges and stripes.  
{\small$L_P$} is a regularization item based on the parameters of operations. {\small$p_o$} denotes the number of parameters of operation $o$. {\small$L_P$} utilizes the number of the parameters as the weight of architecture parameter $\alpha$, so as to reduce the {\small$\alpha$} of the operations which have a large number of parameters and push the algorithm to search for lightweight operations. 
The {\small$\mu$} and {\small$\gamma$} are weighting parameters for balancing the reconstruction performance and model complexity, respectively.  
As for retraining the searched networks, the last term {\small$L(\theta )$} in the total loss function \eqref{eq:loss_function} is removed by setting {\small$\gamma = 0$}.

\begin{figure*}
\begin{center}
\subfloat[The searched network structure of DLSR]{\includegraphics[width=0.45\textwidth]{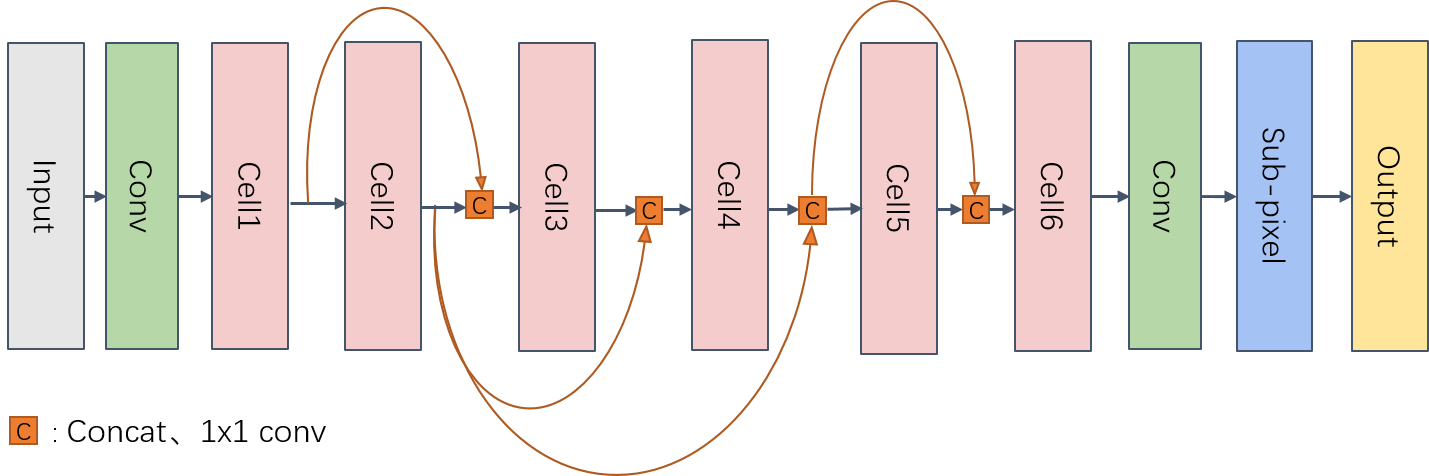}}\!\!\!\!\!\!\!
\hspace{1cm}
\subfloat[The searched cell structure of DLSR]{\includegraphics[width=0.4\textwidth]{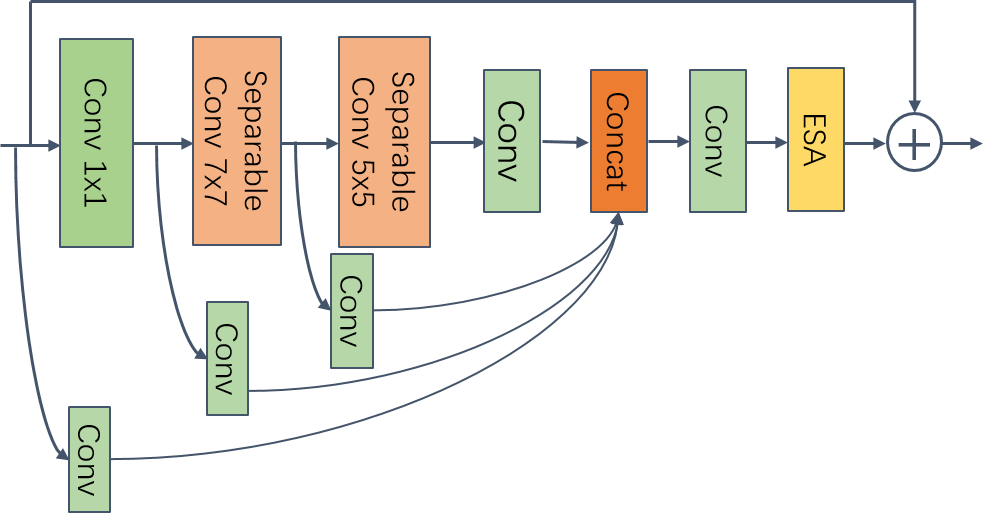}}\!\!\!\!\!\!\!
\hspace{1cm}
\subfloat[The searched network structure of DLSR-DIV]{\includegraphics[width=0.45\textwidth]{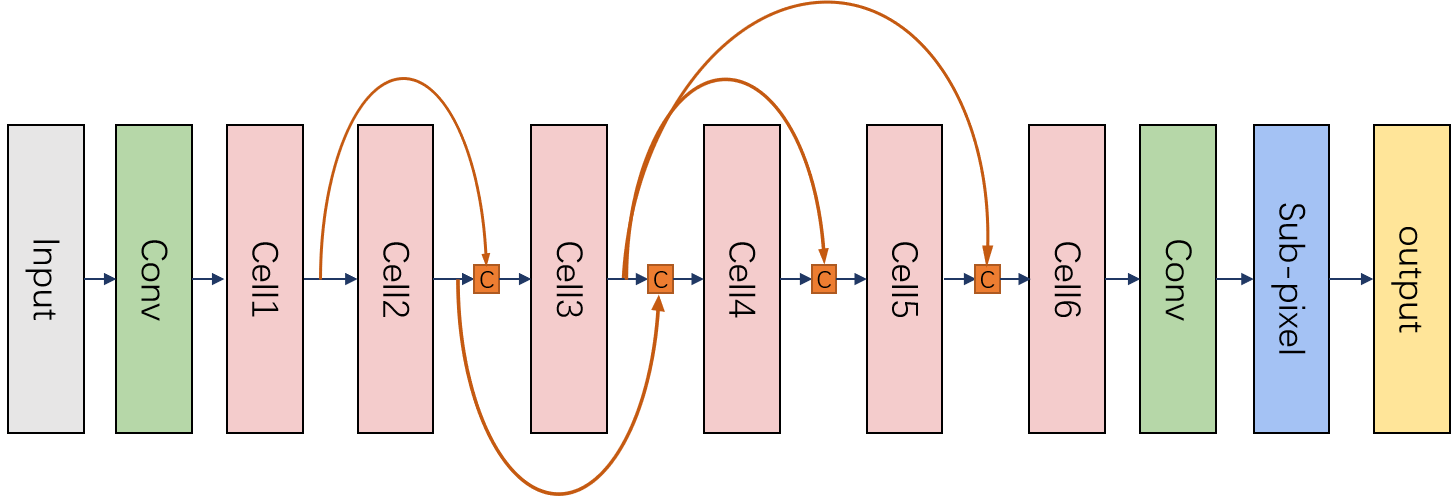}}\!\!\!\!\!\!\!
\hspace{1cm}
\subfloat[The searched cell structure of DLSR-DIV]{\includegraphics[width=0.4\textwidth]{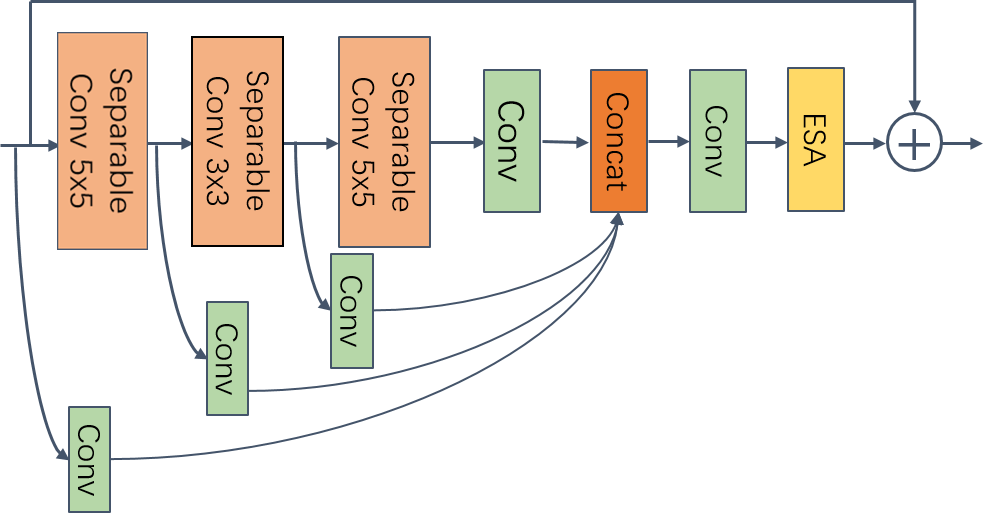}}
\end{center}
\vspace{-0.1cm}
   \caption{The searched network and cell structures. The connections from each cell to the last convolution layer is omitted for clarity.}
\label{fig:beta21}
\vspace{-0.1cm}
\end{figure*}

\SetKwInput{KwInput}{Input}
\SetKwInput{KwOutput}{Output}
\begin{algorithm}[ht]  
  \caption{Searching and training Algorithm} 
  \label{alg:overall}
  \KwInput{Training set {\small$\mathbb{D}$}}  
  Initialize the super-network {\small$\mathcal{T}$} with architecture parameters {\small$\alpha$} and {\small$\beta$}.\\
  Split training set {\small$\mathbb{D}$} into {\small$\mathbb{D}_{train}$} and {\small$\mathbb{D}_{valid}$}.\\
  Train the super-network {\small$\mathcal{T}$} on {\small$\mathbb{D}_{train}$} for several steps to warm up.\\
  \For{{\small$t = 1,2,\ldots, T$}}
  {Sample train batch {\small$\mathbb{B}_{t} = \left \{ {(x_{i}, y_{i})} \right \}_{i=1}^{batch}$} from {\small$\mathbb{D}_{train}$}\\
  Optimize {\small$\theta$} on the {\small$\mathbb{B}_{t}$} by Eq. \eqref{eq:theta_update} \\
  Sample valid batch {\small$\mathbb{B}_{v} = \left \{ {(x_{i}, y_{i})} \right \}_{i=1}^{batch}$} from {\small$\mathbb{D}_{valid}$}\\
  Optimize {\small$\alpha$} and {\small$\beta$} on the {\small$\mathbb{B}_{v}$} by Eq. \eqref{eq:A_update} \\
  Save the genotypes of the searched networks
  }
  Train searched networks from the scratch  \\
  Pick up the best performing network {\small$\mathcal{S}$} \\
  \KwOutput{A lightweight SR network {\small$\mathcal{S}$}}  
\end{algorithm}

%% file: 5.Experiment.tex

\section{Experiments}
\subsection{Datasets}
We use high-quality DIV2K \cite{agustsson2017ntire} and Flickr2K \cite{timofte2017ntire} datasets as training datasets. The DIV2K dataset consists of 800 training images and the Flicker2K dataset consists of 2650 training images. The LR images are obtained by the bicubic downsampling of HR images. In addition, we use the standard benchmark datasets, Set5 \cite{bevilacqua2012low}, Set14 \cite{yang2010image}, B100 \cite{martin2001database}, and Urban100 \cite{huang2015single} as test datasets.

\begin{table*}
\begin{small}
\caption{Image super-resolution results with scale factors of 2, 3, 4 on benchmark datasets.}
\vspace{-0.1cm}
\label{tab:benchmark_result}
\begin{center}
\begin{tabular}{|l|c|c|c|c|c|c|c|}
\hline
\multirow{2}{7em}{Method} & \multirow{2}{2em}{Scale} & Params & Multi-Adds& Set5 & Set14 & B100 & Urban100 \\
\cline{5-8}
& &(K) &(G) & PSNR/SSIM & PSNR/SSIM & PSNR/SSIM & PSNR/SSIM \\
\hline\hline
Bicubic & \multirow{9}{2em}{$\times 2$} & - & - & 33.66/0.9299&30.24/0.8688&29.56/0.8403&26.88/0.8403\\
DRRN \cite{tai2017image} &  & 297 & 6,796.9&37.74/0.9591&33.23/0.9136&32.05/0.8973&31.23/0.9188\\
CARN-M \cite{ahn2018fast} &  & 412 & 91.2& 37.53/0.9583&33.26/0.9141&31.92/0.8960&31.23/0.9194\\
FALSR-B \cite{9413080} &  & 326 & 74.7 &37.61/0.9585&33.29/0.9143&31.97/0.8967&31.28/0.9191\\
ESRN-V \cite{song2020efficient} &  & 324 & 73.4 &37.85/0.9600&33.42/0.9161&32.10/0.8987&31.79/0.9248\\
IMDN \cite{Hui-IMDN-2019} & & 694 & - & 38.00/0.9605&33.63/0.9177&32.19/0.8996&32.17/0.9283\\
PAN \cite{10.1007/978-3-030-67070-2_3} &  & \bf{261} & 70.5 &38.00/0.9605&33.59/0.9181&32.18/0.8997&32.01/0.9273\\
RFDN \cite{liu2020residual} &  & 534 & 123.0 & 38.05/0.9606&\bf{33.68/0.9184}&32.16/0.8994&32.12/0.9278\\
DLSR-DIV &  & 328 & 69.3 & \bf{38.05/0.9607}&33.67/0.9181&32.21/0.8998&32.22/0.9292\\
\bf{DLSR(Ours)} &  & 322 & \bf{68.1} &38.04/0.9606&33.67/0.9183&\bf{32.21/0.9002}&\bf{32.26/0.9297}\\
\hline
\hline
Bicubic & \multirow{8}{2em}{$\times 3$} & - & - &30.39/0.8682&27.55/0.7742&27.21/0.7385&24.46/0.7349\\
DRRN \cite{tai2017image} &  & 297 & 6,796.9 &34.03/0.9244&29.96/0.8349&28.95/0.8004&27.53/0.8378\\
CARN-M \cite{ahn2018fast}&  & 412 & 46.1 &33.99/0.9236&30.08/0.8367&28.91/0.8000&27.55/0.8385\\
ESRN-V \cite{song2020efficient} &  & 324 & 36.2 &34.23/0.9262&30.27/0.8400&29.03/0.8039&27.95/0.8481\\
IMDN \cite{Hui-IMDN-2019} &  & 703 & -  &34.36/0.9270&30.32/0.8417&29.09/0.8046&28.17/0.8519\\
PAN \cite{10.1007/978-3-030-67070-2_3}&  & \bf{261} & 39.0 &34.40/0.9271&30.36/0.8423&29.11/0.8050&28.11/0.8511\\
RFDN \cite{liu2020residual}&  & 541 & 55.4 &34.41/0.9273&30.34/0.8420&29.09/0.8050&28.21/0.8525\\
DLSR-DIV& & 334& 31.5&34.47/0.9277&\bf{30.41/0.8430}&29.12/0.8053&28.25/0.8541\\
\bf{DLSR(Ours)} & & 329 & \bf{30.9} &\bf{34.49/0.9279
}&30.39/0.8428&\bf{29.13/0.8061}&\bf{28.26/0.8548}\\
\hline
\hline
Bicubic & \multirow{8}{2em}{$\times 4$} & - & - &28.42/0.8104&26.00/0.7027&25.96/0.6675&23.14/0.6577\\
DRRN \cite{tai2017image} &  & 297 & 6,796.9 &31.68/0.8888&28.21/0.7720&27.38/0.7284&25.44/0.7638\\
CARN-M \cite{ahn2018fast}&  & 412 & 32.5 &31.92/0.8903&28.42/0.7762&27.44/0.7304&25.62/0.7694\\
ESRN-V \cite{song2020efficient}&  & 324 & 20.7 &31.99/0.8919&28.49/0.7779&27.50/0.7331&25.87/0.7782\\
IMDN \cite{Hui-IMDN-2019}& & 715 & - &32.21/0.8948&28.58/0.7811&27.56/0.7353&26.04/0.7838\\
PAN \cite{10.1007/978-3-030-67070-2_3}&  & \bf{272} & 28.2 &32.13/0.8948&28.61/0.7822&27.59/0.7363&26.11/0.7854\\
RFDN \cite{liu2020residual}&  & 550 & 31.6 &32.24/0.8952&28.61/0.7819&27.57/0.7360&26.11/0.7858\\
DLSR-DIV& &343 & 18.2&32.25/0.8952&28.61/0.7814&27.58/0.7365&26.13/0.7873\\
\bf{DLSR(Ours)} & & 338 & \bf{17.9} &\bf{32.33/0.8963
}&\bf{28.68/0.7832}&\bf{27.61/0.7374}&\bf{26.19/0.7892}\\
\hline
\end{tabular}
\end{center}
\end{small}
\vspace{-0.3cm}
\end{table*}

\begin{table*}[ht]
\begin{small}
\begin{center}
\caption{Comparison results with TPSR-NOGAN on benchmark datasets.}
\label{tab:TPSR_comparison}
\begin{tabular}{|l|c|c|c|c|c|c|c|}
\hline
\multirow{2}{7em}{Method} & \multirow{2}{2em}{Scale} & Params & Multi-Adds& Set5 & Set14 & B100 & Urban100 \\
\cline{5-8}
& &(K) &(G) & PSNR/SSIM & PSNR/SSIM & PSNR/SSIM & PSNR/SSIM \\
\hline\hline
TPSR-NOGAN & $\times 2$ & 60 & 14.0 &37.38/0.9583&33.00/0.9123&31.75/0.8942&30.61/0.9119\\
\bf{DLSR-S(Ours)} &$\times 2$ & \bf{56} & \bf{12.4} & \bf{37.71/0.9595}&\bf{33.33/0.9150}&\bf{31.96/0.8973}&\bf{31.26/0.9196}\\
\hline
\hline
TPSR-NOGAN & $\times 4$ & \bf{61} & 3.6 &31.10/0.8779&27.95/0.7663&27.15/0.7214&24.97/0.7456\\
\bf{DLSR-S(Ours)} &$\times 4$ & 62 & \bf{3.4} & \bf{31.75/0.8885}&\bf{28.31/0.7745}&\bf{27.38/0.7298}&\bf{25.47/0.7663}\\
\hline
\end{tabular}
\end{center}
\end{small}
\vspace{-0.5cm}
\end{table*}

\begin{table}[t]
\vspace{-0.3cm}
\begin{small}
\caption{Efficiency comparison on DIV2K validation set for x4 upscaling}
\begin{center}
\begin{tabular}{|l|p{2.5em}<{\centering}|p{2.5em}<{\centering}|p{2.5em}<{\centering}|p{2.4em}<{\centering}|p{2.7em}<{\centering}|}
\hline
\multirow{2}{1.5em}{Method}& \!\!\!Val & Time & Params & FLOPs & Mem  \\
& PSNR & (ms) & (M) & (G) & (M)\\
\hline
RFDN\cite{liu2020residual} & 29.04 & 98.66 & 0.433 & 27.10 & 788.13 \\
IMDN\cite{Hui-IMDN-2019} &29.13 & 102.12 & 0.894 & 58.53 & 468.29 \\
FMEN-S\cite{du2022fast} &29.00 & 85.85 & 0.341 & 22.28 & 306.74 \\
BSRN-S\cite{li2022blueprint} &29.01 & 95.47 & 0.156 & 9.50 & 730.09 \\
\bf{DLSR(Ours)} &29.05 & 101.38 & 0.339 & 20.41 & 853.65 \\
\hline
\end{tabular}
\label{tab:Efficiency comparison}
\end{center}
\end{small}
\vspace{-0.3cm}
\end{table}


\begin{figure*}[ht]
\centering

\subfloat[HR]{
\begin{minipage}[t]{0.15\linewidth}
\centering
\label{Fig.x2_hr}
\includegraphics[width=1.1in]{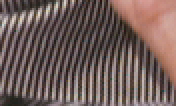}\\
\includegraphics[width=1.1in]{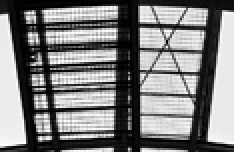}\\
\includegraphics[width=1.1in]{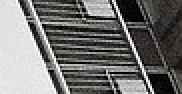}
\end{minipage}
}
\subfloat[Bicubic]{
\begin{minipage}[t]{0.15\linewidth}
\centering
\label{Fig.x2_bic}
\includegraphics[width=1.1in]{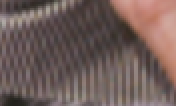}\\
\includegraphics[width=1.1in]{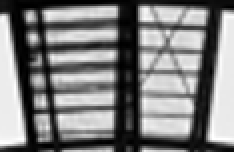}\\
\includegraphics[width=1.1in]{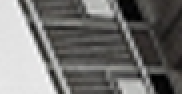}
\end{minipage}
}
\subfloat[CARN-M]{
\begin{minipage}[t]{0.15\linewidth}
\centering
\label{Fig.x2_carn-m}
\includegraphics[width=1.1in]{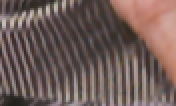}\\
\includegraphics[width=1.1in]{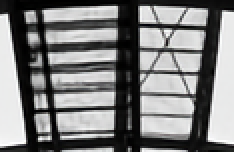}\\
\includegraphics[width=1.1in]{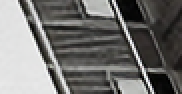}
\end{minipage}
}
\subfloat[FALSR-B]{
\begin{minipage}[t]{0.15\linewidth}
\centering
\label{Fig.x2_FALSR-B}
\includegraphics[width=1.1in]{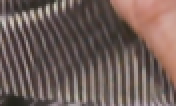}\\
\includegraphics[width=1.1in]{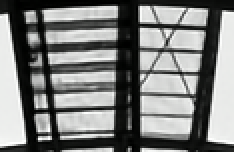}\\
\includegraphics[width=1.1in]{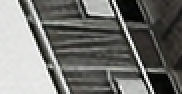}
\end{minipage}
}
\subfloat[PAN]{
\begin{minipage}[t]{0.15\linewidth}
\centering
\label{Fig.x2_PAN}
\includegraphics[width=1.1in]{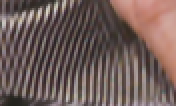}\\
\includegraphics[width=1.1in]{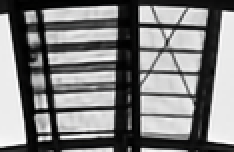}\\
\includegraphics[width=1.1in]{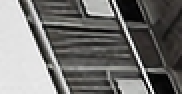}
\end{minipage}
}
\subfloat[Ours]{
\begin{minipage}[t]{0.15\linewidth}
\centering
\label{Fig.x2_set14_OURS}
\includegraphics[width=1.1in]{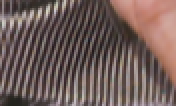}\\
\includegraphics[width=1.1in]{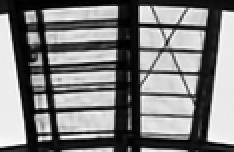}\\
\includegraphics[width=1.1in]{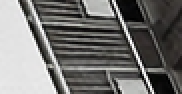}
\end{minipage}
}
\caption{Visual comparisons among SOTA lightweight models in $\times 2$ image super-resolution. The test image patches are from Set14 and Urban100. Note that the results of FALSR-B are based on our test with the pre-trained model which is released by the authors. The results of CARN-M and PAN are directly taken from the authors' release. Our method has better reconstruction performance on image details, such as thin stripes on the clothes and edges of windows.}
\label{Fig.main}
\vspace{-0.2cm}
\subfloat[HR]{
\begin{minipage}[t]{0.15\linewidth}
\centering
\label{Fig.x4_hr}
\includegraphics[width=1.1in]{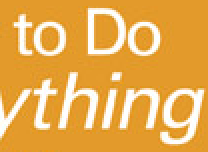}\\
\includegraphics[width=1.1in]{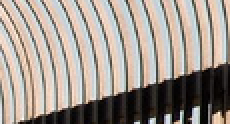}\\
\includegraphics[width=1.1in]{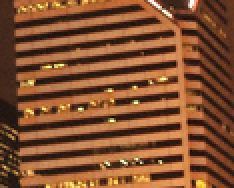}\\
\includegraphics[width=1.1in,height=1.2in]{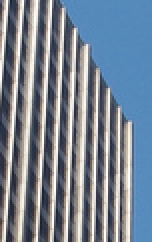}
\end{minipage}
}
\subfloat[Bicubic]{
\begin{minipage}[t]{0.15\linewidth}
\centering
\label{Fig.x4_bic}
\includegraphics[width=1.1in]{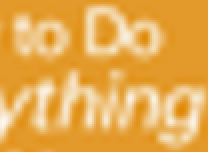}\\
\includegraphics[width=1.1in]{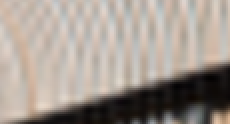}\\
\includegraphics[width=1.1in]{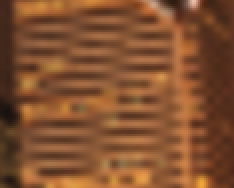}\\
\includegraphics[width=1.1in,height=1.2in]{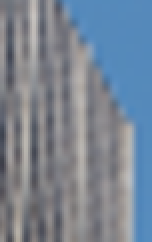}
\end{minipage}
}
\subfloat[CARN-M]{
\begin{minipage}[t]{0.15\linewidth}
\centering
\label{Fig.x4_carn-m}
\includegraphics[width=1.1in]{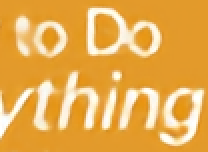}\\
\includegraphics[width=1.1in]{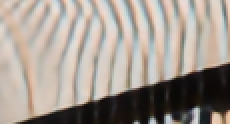}\\
\includegraphics[width=1.1in]{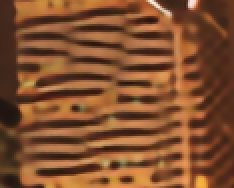}\\
\includegraphics[width=1.1in,height=1.2in]{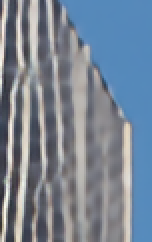}
\end{minipage}
}
\subfloat[RFDN]{
\begin{minipage}[t]{0.15\linewidth}
\centering
\label{Fig.x4_RFDN}
\includegraphics[width=1.1in]{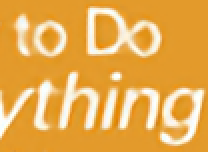}\\
\includegraphics[width=1.1in]{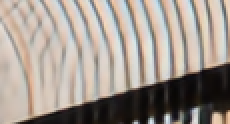}\\
\includegraphics[width=1.1in]{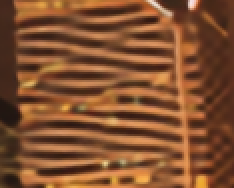}\\
\includegraphics[width=1.1in,height=1.2in]{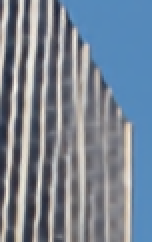}
\end{minipage}
}
\subfloat[PAN]{
\begin{minipage}[t]{0.15\linewidth}
\centering
\label{Fig.x4_PAN}
\includegraphics[width=1.1in]{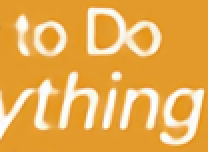}\\
\includegraphics[width=1.1in]{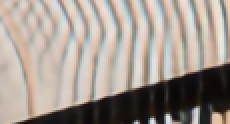}\\
\includegraphics[width=1.1in]{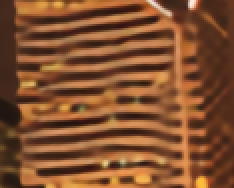}\\
\includegraphics[width=1.1in,height=1.2in]{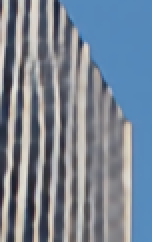}
\end{minipage}
}
\subfloat[Ours]{
\begin{minipage}[t]{0.15\linewidth}
\centering
\label{Fig.x4_OURS}
\includegraphics[width=1.1in]{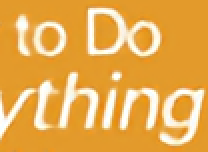}\\
\includegraphics[width=1.1in]{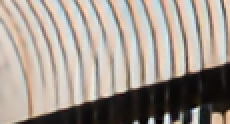}\\
\includegraphics[width=1.1in]{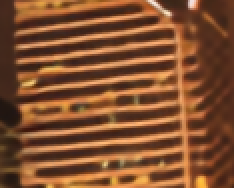}\\
\includegraphics[width=1.1in,height=1.2in]{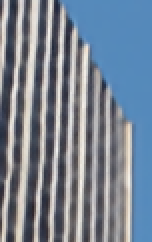}
\end{minipage}
}
\caption{Visual comparisons among SOTA lightweight models in $\times 4$ image super-resolution. The test image patches are from Set14 and Urban100. Note that the results of RFDN are based on our test with the pre-trained model which is officially released by the authors. Our method shows better reconstruction performance and less deformation on image details such as texts and stripes.}
\label{Fig.main_X4}
\vspace{-0.6cm}
\end{figure*}

\subsection{Implementation Details}
We merge the DIV2K and Flickr2K datasets together and denote them as DF2K dataset with a total of 3450 images.
During the searching stage, we split the dataset to 3000 images as training dataset  {\small$\mathbb{D}_{train}$} and the remaining 450 images as validation dataset {\small$\mathbb{D}_{valid}$}. We augment the datasets by random rotations of $90^\circ$, $180^\circ$, $270^\circ$, and horizontal flips. The high-resolution (HR) patch size is set as 64 and the minibatch size is set as 64. We optimize the $\theta$, $\alpha$ and $\beta$ parameters with ADAM optimizer \cite{kingma2014adam} with $2 \times 10^5$ iterations.
For parameter $\theta$, the learning rate is set to $3 \times 10^{-4}$, the momentum parameter and exponential moving average parameter are set as (0.9,0.999) and the weight decay is set to $10^{-8}$.
For parameters $\alpha$ and $\beta$, the learning rate is set to $3 \times 10^{-4}$,  the momentum parameter and exponential moving average parameter are set as (0.5,0.999) and the weight decay is set to $10^{-8}$.
The warm-up process takes $2 \times 10^4$ steps that only parameter $\theta$ is updated. The learning rates of the warm-up process and searching process are both set to $3 \times 10^{-4}$.
We save the genotypes of the searched models at about $5 \times 10^4$ step, $10^5$ step, and $1.5 \times 10^5$ step when the distribution of the architecture parameters turn stable during searching.
The number of channels is set to 48 and the number of cells is set to 6. The hyper-parameter $\lambda$ is set as 1.0, $\mu$ is set as 0.2, and $\gamma$ is set as 0.2.

For retraining the searched networks, we use the whole DF2K dataset with the same data augmentation as the searching stage. For $\times 2, \times 3, \times 4 $ super-resolution, HR patch size is set as 128, 192, and 256, respectively.
We train our searched DLSR model with ADAM optimizer \cite{kingma2014adam} with the same settings as the optimization of the parameter $\theta$ during the searching stage. We train the model in $2 \times 10^6$ steps and set the minibatch size as 32.
The learning rate is initialized with $3 \times 10^{-4}$ and halved every $4 \times 10^5$ steps.
The weights of both $\times 3, \times 4 $ super-resolution models are warmed up by the weight of the pre-trained $\times 2$ SR model. 
We perform $\times 2$ SR for searching the neural network architectures and apply the searched models to $\times2, \times 3, \times 4$ SR tasks. Both the searching and training stages are performed on a single NVIDIA Tesla V100 GPU. All the experiments are conducted in PyTorch 1.2 and Python 3.7.

\subsection{Searched Results}
The searched network structure and cell structure are shown in Figure \ref{fig:beta21}. For clarity, we omit the connections between each block and the end of the model in the figure. The searched cell is made up of a $1\! \times\! 1$ convolution layer, $7\! \times\! 7$ separable convolution layer, $5 \times 5$ separable convolution layer, ESA block, and residual connections with information distillation mechanism. Since the parameters and FLOPS of the $1 \!\times\! 1$ convolution, $5 \!\times\! 5$ separable convolution, and $7\!\times\!7$ separable convolution are all fewer than the original $3\!\times\!3$ convolution, we obtain a much smaller (nearly half the original size) model compared with vanilla RFDN \cite{10.1007/978-3-030-67070-2_2}.

\subsection{Comparison with State-of-the-art Methods}
We compare the DLSR model with state-of-the-art lightweight SR methods on two commonly-used metrics: peak signal-to-noise ratio (PSNR) and structural similarity index measure (SSIM) \cite{wang2004image} on the Y channel of the transformed YCbCr space. We also present the number of the parameters and number of the operations (Multi-Adds) to show the model complexity. Multi-Adds is calculated on 720p ($1280 \times 720$) HR images. The $\times2, \times 3, \times 4 $ SR results are shown in Table \ref{tab:benchmark_result}, and the best results are highlighted. The DLSR-DIV denotes that the model is trained only with DIV2K dataset for fair comparisons. The network and cell structures of DLSR-DIV are shown in Figure \ref{fig:beta21} (c) and (d)
The visual performance of $\times2,\times4$ super-resolution are shown in Figures \ref{Fig.main} and \ref{Fig.main_X4}. Compared with DRRN \cite{tai2017image}, our method only takes $1\%$ Multi-Adds, while achieving 1dB PSNR improvement on the Urban100 dataset in $\times 2, \times 3$ SR tasks, 0.7dB PSNR improvement on the Set5 dataset in $\times 4$ SR task, as well as 0.3-0.7dB PSNR improvement on other tasks, respectively. Our method surpasses other NAS based methods like FALSR-B \cite{9413080} and ESRN-V \cite{song2020efficient} by a large margin with 0.2-0.4dB PSNR improvement with fewer parameters and Multi-Adds in most of the SR tasks (Table \ref{tab:benchmark_result}).
As shown in Table \ref{tab:GPU_days}, the search cost of our method is significantly less than NAS-based SR methods. FALSR-B \cite{9413080} takes less than 3 days on 8 GPUs to execute their pipeline once. ESRN-V \cite{song2020efficient} takes around one day on 8 GPUs to execute their evolution procedure. Our method only takes around 2 days on one GPU.

Compared with hand-crafted light-weight models like IMDN \cite{Hui-IMDN-2019} and RFDN \cite{10.1007/978-3-030-67070-2_2}, the DLSR method only takes about half the amount of parameters while still outperforming them. Compared with PAN \cite{10.1007/978-3-030-67070-2_3}, which is the most lightweight deep SR model in AIM2020 Efficient Super Resolution, our method is still able to outperform it with fewer Multi-Adds.

The Efficiency Comparison results in terms of inference time, parameters, FLOPs, and Memory are listed in \ref{tab:Efficiency comparison}. The experiments are with PyTorch 1.10.0, CUDA Toolkit 11.6, and cuDNN 8.3.0.2, on
an NVIDIA 3080 GPU. In addition to the baseline methods like IMDN and RFDN, the winners of the NTIRE 2022 Challenge on Efficient Super Resolution are also included. Please note that our method is not specially optimized for the challenge. Our method achieves better PSNR results with comparable parameters and FLOPs. Since it is unfriendly for GPU to calculate separable convolution, the running time of our method is relatively larger. Because our method introduces the connections of the network-level, the memory usage is larger for preserving the output features.

{
\begin{table*}
\begin{small}
\begin{center}
\caption{Comparison results between the model with/without network-level connections DLSR/DLSR-B.}
\label{tab:DLSR-B_comparison}
\begin{tabular}{|l|c|c|c|c|c|}
\hline
\multirow{2}{2em}{Method} & \multirow{2}{2em}{Scale}&Set5&Set14&B100&Urban100\\
\cline{3-6}
 & & PSNR/SSIM & PSNR/SSIM & PSNR/SSIM & PSNR/SSIM \\
\hline
DLSR-B & $\times 2$ & 38.04/0.9606 & 33.63/0.9177 &32.20/0.9000 & 32.20/0.9293\\
DLSR & $\times 2$ & 38.04/0.9606&\bf{33.67}/0.9183&\bf{32.21/0.9002}&\bf{32.26}/0.9297\\
\hline
\hline
DLSR-B & $\times 4$ & 32.27/0.8959 & 28.67/0.7832 &27.60/0.7372 & 26.16/0.7885\\
DLSR & $\times 4$ & \bf{32.33}/0.8963&\bf{28.68}/0.7832&\bf{27.61}/0.7374&\bf{26.19}/0.7892\\
\hline
\end{tabular}
\end{center}
\end{small}
\vspace{-0.1cm}
\begin{small}
\begin{center}
\caption{Comparison results with different loss function configurations on benchmark datasets.}
\label{tab:loss_comparison}
\begin{tabular}[h]{|l|c|c|c|c|c|c|c|}
\hline
\multirow{2}{7em}{Method} & \multirow{2}{2em}{Scale} & Params & Multi-Adds& Set5 & Set14 & B100 & Urban100 \\
\cline{5-8}
& &(K) &(G) & PSNR/SSIM & PSNR/SSIM & PSNR/SSIM & PSNR/SSIM \\
\hline\hline
DLSR-L1 & $\times 2$ & 323 & \bf{68.1} &\bf{38.04/0.9606}&33.69/0.9185&32.20/0.9002&32.27/0.9297\\
DLSR-HFEN &$\times 2$ & 365 & 77.8 & 38.02/0.9605&\bf{33.78/0.9200}&\bf{32.21/0.9003}&\bf{32.34/0.9305}\\
DLSR & $\times 2$ & \bf{322} & \bf{68.1} & \bf{38.04/0.9606}&33.67/0.9183&{\bf32.21}/0.9002&32.26/0.9297\\
\hline
\end{tabular}
\end{center}
\end{small}

\end{table*}
}

\begin{table}[!t]
\vspace{-0.2cm}
\begin{small}
\caption{Searching cost of NAS based SR methods}
\begin{center}
\begin{tabular}{|c|c|}
\hline
NAS based SR method & GPU days  \\
\hline
FALSR \cite{9413080} & 24 \\
ESRN \cite{song2020efficient} & 8 \\
\bf{DLSR(ours)}& \bf{2} \\
\hline
\end{tabular}
\label{tab:GPU_days}
\end{center}
\end{small}
\vspace{-0.7cm}
\end{table}

\begin{figure*}[ht]
\centering
\subfloat[HR]{
\begin{minipage}[t]{0.18\linewidth}
\centering
\label{Fig.x2_HR}
\includegraphics[width=1.2in]{comparison_result/x2/Set14/HR}\\
\includegraphics[width=1.2in]{comparison_result/x2/Urban100/img_092/HR}\\
\end{minipage}
}
\subfloat[DLSR-L1]{
\begin{minipage}[t]{0.18\linewidth}
\centering
\label{Fig.x2_l1}
\includegraphics[width=1.2in]{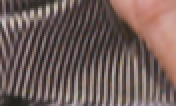}\\
\includegraphics[width=1.2in]{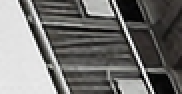}\\
\end{minipage}
}
\subfloat[DLSR-HFEN]{
\begin{minipage}[t]{0.18\linewidth}
\centering
\label{Fig.x2_HFEN}
\includegraphics[width=1.2in]{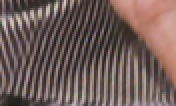}\\
\includegraphics[width=1.2in]{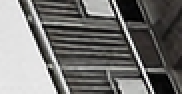}\\
\end{minipage}
}
\subfloat[DLSR]{
\begin{minipage}[t]{0.18\linewidth}
\centering
\label{Fig.x2_DLSR}
\includegraphics[width=1.2in]{comparison_result/x2/Set14/Ours}\\
\includegraphics[width=1.2in]{comparison_result/x2/Urban100/img_092/OURS}\\
\end{minipage}
}
\vspace{-0.05cm}
\caption{Visual comparisons among the models trained with different loss configurations in $\times 2$ image super-resolution. DLSR-L1 model is searched and retrained only with L1 loss. DLSR-HFEN model is searched and retrained with L1 loss and HFEN loss. DLSR model is searched and retrained with L1 loss, HFEN loss, and parameter regularization.}
\label{Fig.comparision_loss}
\vspace{-0.1cm}
\end{figure*}

The visual comparison results show that our method achieves better performance in reconstructing image details such as tiny stripes and the edges of the text. In Figure \ref{Fig.main}, our DLSR method reconstructs the right direction of the thin stripes on the clothes with high visual quality and successfully reconstructs the edge of the window, while other methods cannot. In Figure \ref{Fig.main_X4}, our method reconstructs the round edge of the text 'O' and other stripe-like image details without distortion. In summary, the quantitative and visual results both demonstrate that our models outperform the state-of-the-art SR models on multiple datasets and scales with fewer parameters and Multi-Adds.

In addition, to compare our method with the Tiny Perceptual Super Resolution (TPSR) model \cite{lee2020journey}, which is a super lightweight SR model with 60K parameters, we cut the channel number of our DLSR model to 18 and denote the smaller model as DLSR-S. The $\times 2, \times 4$ SR comparison result is shown in Table \ref{tab:TPSR_comparison}. As our method is not based on the generative adversarial networks (GAN), we compare it with the baseline model of the TPSR method called TPSR-NOGAN. The result indicates that even with fewer Multi-Adds, our DLSR-S model can still surpass the TPSR-NOGAN with a large margin of PSNR and SSIM. 

\begin{table}[ht]
\footnotesize
\footnotesize
\begin{center}
\caption{validation of the effectiveness of cell-level search space.}
\vspace{-0.2cm}
\label{tab:cell_level_comparison}
\begin{tabular}{|p{2.6em}|p{1.5em}<{\centering}|p{2.0em}<{\centering}|p{2.2em}<{\centering}|p{2em}<{\centering}|p{2em}<{\centering}|p{2em}<{\centering}|p{3.3em}<{\centering}|}
\hline
\multirow{2}{0em}{Method\!\!\!\!\!} & \multirow{2}{1.5em}{\!\!\!Scale\!\!} & \!\!\!Params & \!\!FLOPs& Set5 & Set14 & B100 & \!\!\!Urban100\!\! \\
\cline{5-8}
& &(K) &(G) & \!\!PSNR & PSNR & PSNR & PSNR \\
\hline\hline
RFDN & $\times 2$ & 534 & 123.0 & \bf{38.05}&\bf{33.68}&32.16&32.12\\
\!\!\!\bf{DLSR-B\!\!\!\!\!} &$\times 2$ & \bf{322} & \bf{68.1} & 38.04 & 33.63 &\bf{32.20} & \bf{32.20}\\
\hline
\hline
RFDN & $\times 4$ & 550 & 31.6 &32.24&28.61&27.57&26.11\\
\!\!\!\bf{DLSR-B\!\!\!\!\!} &$\times 4$ &\bf{338} & \bf{17.9}& \bf{32.27} & \bf{28.67} &\bf{27.60} & \bf{26.16}\\
\hline
\end{tabular}
\end{center}
\vspace{-0.5cm}
\end{table}

\subsection{Ablation Studies}

{\bf The effectiveness of cell-level search.}
The cell-level search in our method is based on information distillation structure which is adopted in hand-crafted RFDB [22] structure. While RFDB only contains the same three 3$\times$3 convolution layers, our method aims to search for the combinations of lightweig-ht operations. So, the comparison results between RFDN and DLSR-B (only considering the cell-level search, without network-level search) validate the effectiveness of cell-level search, shown in Table \ref{tab:cell_level_comparison}.

\noindent

{\bf The effectiveness of network-level connections}. To compare with DLSR, we design our method to search for a baseline model only on the cell-level search space. After searching and retraining, we name this model DLSR-B. Coincidentally, the DLSR-B has the same numbers of parameters and Multi-Adds as DLSR. The comparison result is illustrated in Table \ref{tab:DLSR-B_comparison}. The result shows that the network-level connections can improve the performance of DLSR-B. Thus, the search space of network-level connections which we propose is innovative and meaningful.

\noindent
{\bf The effectiveness of the loss function.} The loss function is comprised of three parts: the L1 loss, the HFEN loss, and the parameter loss. We conduct the experiments on three different models: DLSR-L1, DLSR-HFEN, DLSR. The DLSR-L1 model is searched and retrained only with L1 loss. The DLSR-HFEN model is searched and retrained with L1 loss and HFEN loss. The DLSR model is searched and retrained on all three parts of the loss. The comparison result is illustrated in Table \ref{tab:loss_comparison} and Figure \ref{Fig.comparision_loss}. The result shows that the DLSR-HFEN achieves better results than DLSR-L1 with a larger number of parameters and Multi-Adds. Although DLSR achieves similar numerical results comparing with DLSR-L1, the visual results of DLSR are much better than DLSR-L1. In a word, the DLSR model achieves a better trade-off between the SR performance and the model complexity, and the HFEN loss can contribute to a better visual effect with sharper image details. 


\noindent
{\bf The search stability and the  confidence  interval.} To prove of the effectiveness of our proposed DLSR method, we repeat the searching experiment 6 times on $\times2$ SR task and get 6 best-performing sub-networks. The $95\%$ confidence intervals of the results are shown in Table \ref{tab:confidence_intervals}. The results show that our proposed DLSR method is stable and effective.

\begin{table*}
\vspace{-0.3cm}
\begin{small}
\caption{Image super-resolution results of searched models with scale factors of 2 on benchmark datasets.}
\begin{center}
\begin{tabular}{|c|l|c|c|c|c|c|}
\hline
\multirow{2}{3em}{Model}& \multirow{2}{10em}{Cell-level Genotypes} & \!\!\! Params & Set5 & Set14 & B100 & Urban100  \\
\cline{4-7}
& & (K) & PSNR & PSNR & PSNR & PSNR\\
\hline\hline
1 & [conv1x1, sepconv3x3, sepconv7x7] & 322 & 38.04 & 33.67 & 32.21 & 32.26 \\
2 & [sepconv3x3, sepconv3x3, sepconv3x3] & 309 & 38.04 & 33.63 & 32.21 & 32.20 \\
3 & [sepconv5x5, sepconv7x7, sepconv3x3] & 342 & 38.06 & 33.62 & 32.21 & 32.20 \\
4 & [sepconv5x5, sepconv7x7, sepconv7x7] & 365 & 38.03 & 33.74 & 32.23 & 32.33 \\
5 & [dilconv5x5, sepconv3x3, sepconv3x3] & 298 & 38.04 & 33.60 & 32.20 & 32.14 \\

\hline
\end{tabular}
\label{tab: experiment findings}
\end{center}
\end{small}
\vspace{-0.3cm}
\end{table*}

\noindent

{\bf Experimental findings}
 We list some models searched with different random seeds at initialization. The quantitative results are shown in Table \ref{tab: experiment findings}. The cell-level genotypes denote the searched results of MRBs. For simplicity, the network-level structures are not displayed which are also different among the networks.

The results show that when the initialization changes, the searched structures change significantly with different types of convolutional layers, resulting in different performance. As the loss function is design to punish the convolutional layers with large parameters, the conv5x5/conv7x7 hardly ever appear. The structure with dilated convolutional layers is seldom searched and the performance is not the best.
Separatable convolutional layers appear most frequently and achieve better trade-off between lightweight structures and super-resolution performance. 
Moreover, we find that the networks which have the operation ``separate convolution 5$\times$5" achieve better results. At the network-level searching, the information flow directly from the prior cell seems to be the most important, and the features from the second cell are the most frequently adopted. These findings may be instructive for lightweight network design.


\begin{table}[t]
\footnotesize
\begin{center}
\caption{The $95\%$ confidence intervals of $\times2$ SR task on PSNR.}
\label{tab:confidence_intervals}
\begin{tabular}{|l|c|c|c|c|}
\hline
Method & Params & Multi-Adds& Set5 \\

\hline
DLSR  & 335$\pm15.16$& 71.0$\pm3.49$ &38.04$\pm0.008$\\
\hline\hline
Method &Set14 & B100 & Urban100 \\
\hline
DLSR  &33.65$\pm0.032$&32.22$\pm0.007$&32.24$\pm0.039$\\

\hline
\end{tabular}
\end{center}
\vspace{-0.6cm}
\end{table}


%% file: 6.Conclusion.tex
\section{Conclusions}
In this work, we propose a novel {\bf D}ifferentiable neural architecture search approach to search for extremely {\bf L}ightweight single image {\bf S}uper-{\bf R}esolution models on both the cell-level and the network-level, dubbed DLSR. 
In addition, we design a novel loss function that considers distortion, high-frequency reconstruction, and lightweight regularization that jointly pushes the searching direction to explore a better lightweight SR model. 
Experimental results show that our DLSR method can surpass both the hand-crafted and NAS-based SOTA lightweight SR methods in terms of PSNR and SSIM with fewer parameters and Multi-Adds.

\vspace{+1cm}